\documentclass{JINST}

\title{Performance data from the ZEPLIN-III second science run}

\author{P.~Majewski$^a$\thanks{Corresponding author.}, V.N.~Solovov$^b$, D.Yu.~Akimov$^c$, H.M.~Ara\'{u}jo$^d$,
E.J.~Barnes$^e$, V.A.~Belov$^c$, A.A.~Burenkov$^c$, V.~Chepel$^b$, A.~Currie$^d$, L.~DeViveiros$^b$, B.~Edwards$^a$, C.~Ghag$^e$, A.~Hollingsworth$^e$, M.~Horn$^d$, G.E.~Kalmus$^a$, A.S.~Kobyakin$^c$, A.G.~Kovalenko$^c$, V.N.~Lebedenko$^d$, A.~Lindote$^{b,a}$, M.I.~Lopes$^b$, R.~L\"{u}scher$^a$, A.St\,J.~Murphy$^e$, F.~Neves$^{b,d}$, S.M.~Paling$^a$, J.~Pinto da Cunha$^b$, R.~Preece$^a$, J.J.~Quenby$^d$, L.~Reichhart$^e$, P.R.~Scovell$^e$, C.~Silva$^b$, N.J.T.~Smith$^a$, V.N.~Stekhanov$^c$, T.J.~Sumner$^d$, C.~Thorne$^d$, R.J.~Walker$^d$\\
\llap{$^a$}Particle Physics Department, STFC Rutherford Appleton Laboratory, Chilton, UK\\
\llap{$^b$}LIP--Coimbra \& Department of Physics of the University of Coimbra, Portugal\\
\llap{$^c$}Institute for Theoretical and Experimental Physics,Moscow, Russia\\
\llap{$^d$}High Energy Physics group, Blackett Laboratory,Imperial College London, UK\\
\llap{$^e$}School of Physics \& Astronomy, University of Edinburgh, UK\\
E-mail: \email{pawel.majewski@stfc.ac.uk}}

\abstract{ZEPLIN-III is a two-phase xenon direct dark matter experiment located at the 
Boulby Mine (UK). After its first science run in 2008 it was upgraded with: an array of low background 
photomultipliers, a new anti-coincidence detector system with plastic scintillator and an improved
calibration system. After 319 days of data taking the second science run ended in May 2011. In this paper we describe the instrument 
performance with emphasis on the position and energy reconstruction algorithm and summarise the final science results.} 


\begin{document}

\section{Introduction}
The ZEPLIN-III experiment searching for Weakly Interacting Massive Particles (WIMPs) operated in the Palmer Laboratory 1070 m 
underground (2850 m water equivalent) at the Boulby mine (North East of England) between 2006 and 2011. After 3 kg single phase 
ZEPLIN-I~\cite{zep1} and 31 kg double phase ZEPLIN-II~\cite{zep2} running between 2001 and 2006, ZEPLIN-III 
was the third generation of liquid xenon experiments deployed at Boulby. This two-phase xenon detector filled with 12 kg of liquid xenon (LXe) 
detects both scintillation light and ionisation released from particle interactions. The ionisation charge is drifted upward by an applied 
electric field and is emitted into a few mm thick xenon gas layer where it is accelerated 
creating electroluminescence light. Both light signals, S1 and S2, are detected by an array of 31, upward-facing, 2-inch 
photomultipliers located underneath the liquid xenon target. The S2/S1 ratio is used to discriminate electron recoils, caused by
radioactive background, from nuclear recoils caused by WIMP interactions. A detailed description of the detector design can be found in~\cite{zep3}.  

\section{Detector Performance}
After the first science run (FSR) of 83 days~\cite{zep3res1,zep3res2,zep3res3}, the ZEPLIN-III detector was upgraded; this involved 
the replacement of the photomultipliers, inclusion of a new anti-coincidence detector and a new automated calibration source delivery system, as well
as several other calibration-related improvements. The second science run (SSR) started on the 24$^{th}$ June 2010 and ended on the 7$^{th}$ May 2011, delivering 
a raw fiducial exposure of 1343.8 kg$\cdot$days. 

For the SSR new, custom-built low background photmultipliers, model D766Q from ET Enterprises Limited~\cite{etl}, were used reducing the overall background 
gamma radiation in ZEPLIN-III by a factor of 18~\cite{zep3back}. Despite very low radioactivity, 
below 50 mBq/PMT, their optical performance compared to the previous set of PMTs used in the FSR was poor. Their average quantum 
efficiency at 175 nm was 26.2\% instead of 30\% and the gain variation between PMTs was a factor of 100 (max/min), almost 17 times greater than in the FSR. 
This resulted in an energy resolution of 12\% at 122 keV and discrimination power of 280:1, compared to 8.1\% and 7800:1, respectively, in the FSR.

\begin{figure}[ht]
\includegraphics[width=.8\textwidth,angle=-90]{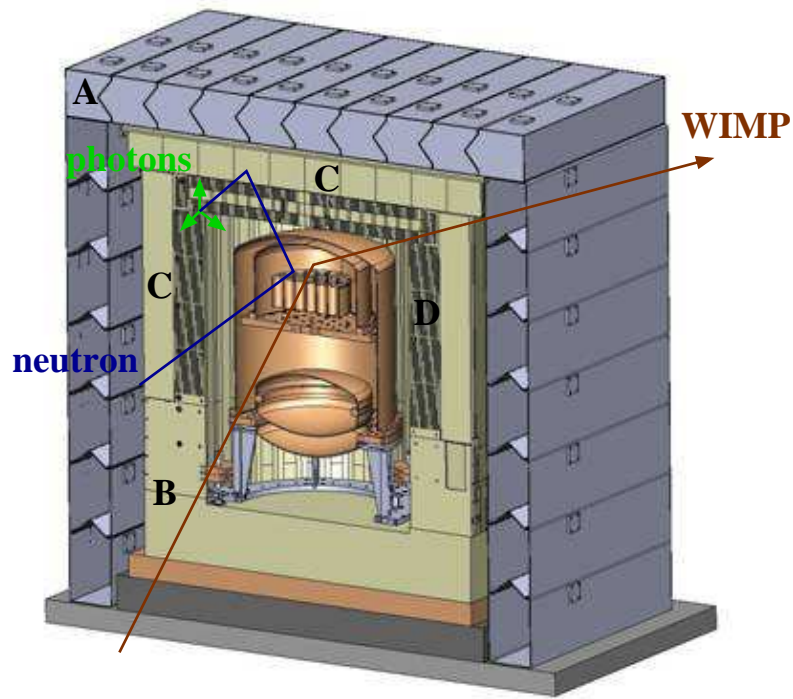}
\caption{Schematic view of the ZEPLIN-III experiment. Copper parts of the ZEPLIN-III detector located in the centre are surrounded by:
the polypropylene structure with no Gd (B), gadolinium-loaded plastic (D), 52 plastic scintillator slabs (C) and lead gamma shield (A).}  
\label{fig1}
\end{figure}

The anti-coincidence system consisted of 32 barrel and 20 roof plastic scintillator slabs coupled to 
a gadolinium-loaded passive polypropylene shield~\cite{vetodes}. As shown in Figure~\ref{fig1}, the veto detector was placed between 
the ZEPLIN-III detector and the lead gamma shield, providing greater than 3$\pi$ sr coverage. Neutrons entering the hydrogen-rich polypropylene 
moderate down to thermal energies and are captured by the $^{157}$Gd.  The neutron cooling and capture process takes an average 10.7 $\mu$s and ends with 
the emission of 3-4 $\gamma$-rays which, with an average total energy of 8 MeV, are detected by the plastic scintillator.

The signal from each scintillator bar was detected by a 3-inch photomultiplier and then digitized with 100 ns sampling rate by a CAEN-1724 ADCs. Waveforms 
of 320 $\mu$s were read out in order to provide information about prompt (PTAG) and delayed (DTAG) coincidences. Calibration with neutron 
sources demonstrated a 60\% efficiency of the neutron tagging. The veto system also provided the rejection of 28\% for $\gamma$-ray background for
prompt coincidences~\cite{vetoperf}. 

\begin{figure}[ht]
\includegraphics[width=.398\textwidth]{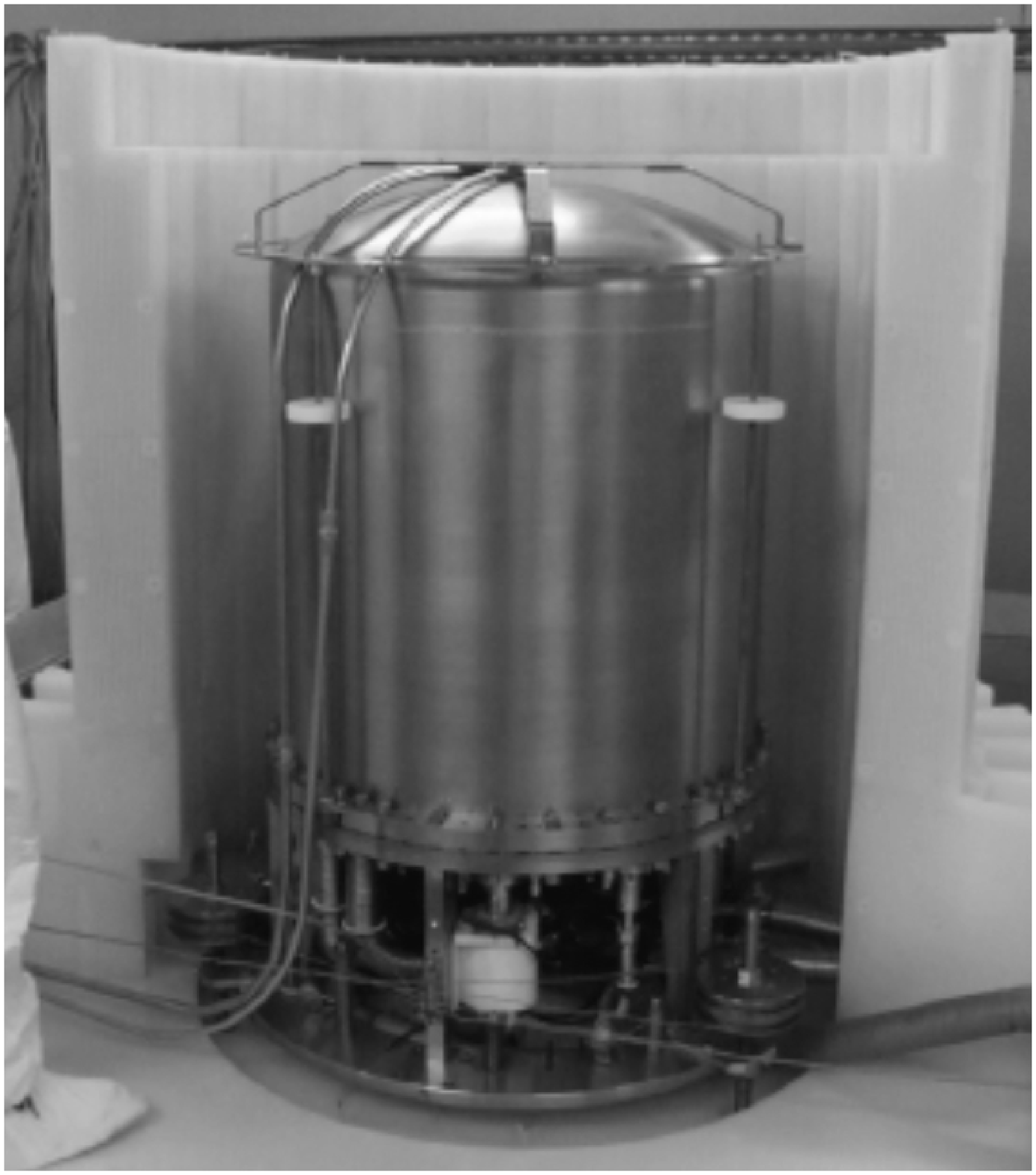}~~~~\includegraphics[width=.602\textwidth]{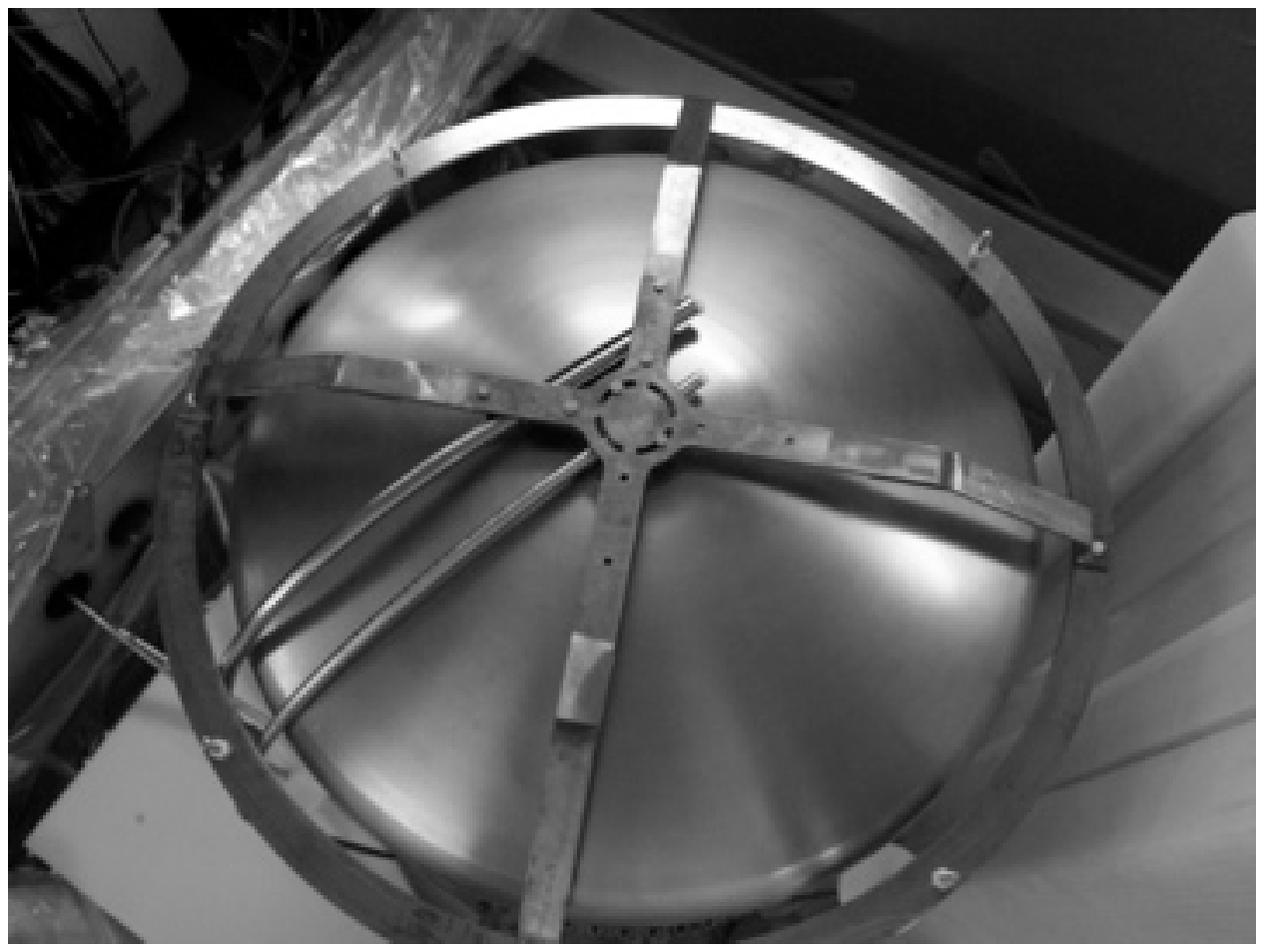}
\caption{Left: ZEPLIN-III partially surrounded by the polypropylene slabs of the veto system. Two pipes of the source delivery system are attached 
to the vacuum dome. The detector leveling system with pulley wheels and cables can be seen at the bottom of the detector. Right: ZEPLIN-III 
viewed from the top. The gamma source was delivered through the pipe above the center of the detector, while the neutron source
was placed on the top 5 centimeters off-centre.}  
\label{fig2}
\end{figure}

During the SSR ZEPLIN-III was equipped with a new automated $\gamma$-ray calibration source delivery system. It consisted of a motorised cable pulley system 
to which a radioactive source was attached, traveling down the pipe connected to the ZEPLIN-III dome as shown in Figure~\ref{fig2}. 

The pipe with a smaller aperture served to guide a $^{57}$Co source for daily $\gamma$-calibration, whereas the wider one guided the manually inserted 
Am-Be source, for the occasional neutron calibrations. The system was driven by the slow-control PC, which was programmed to perform daily calibration. 
The system was completely reliable. 

In addition to the new calibration system a 5.1 mm thick copper `phantom' grid with 3 cm $\times$ 3 cm rectangular voids was placed on top of the anode plate. 
The 122 keV photons from the $^{57}$Co calibration source, located centrally above the LXe target, were attenuated by the phantom grid, creating a shadow 
image on the liquid surface. This was used to test the position reconstruction algorithm and to measure the spatial resolution of the PMT readout. 

\begin{figure}[ht]
\includegraphics[width=.58\textwidth]{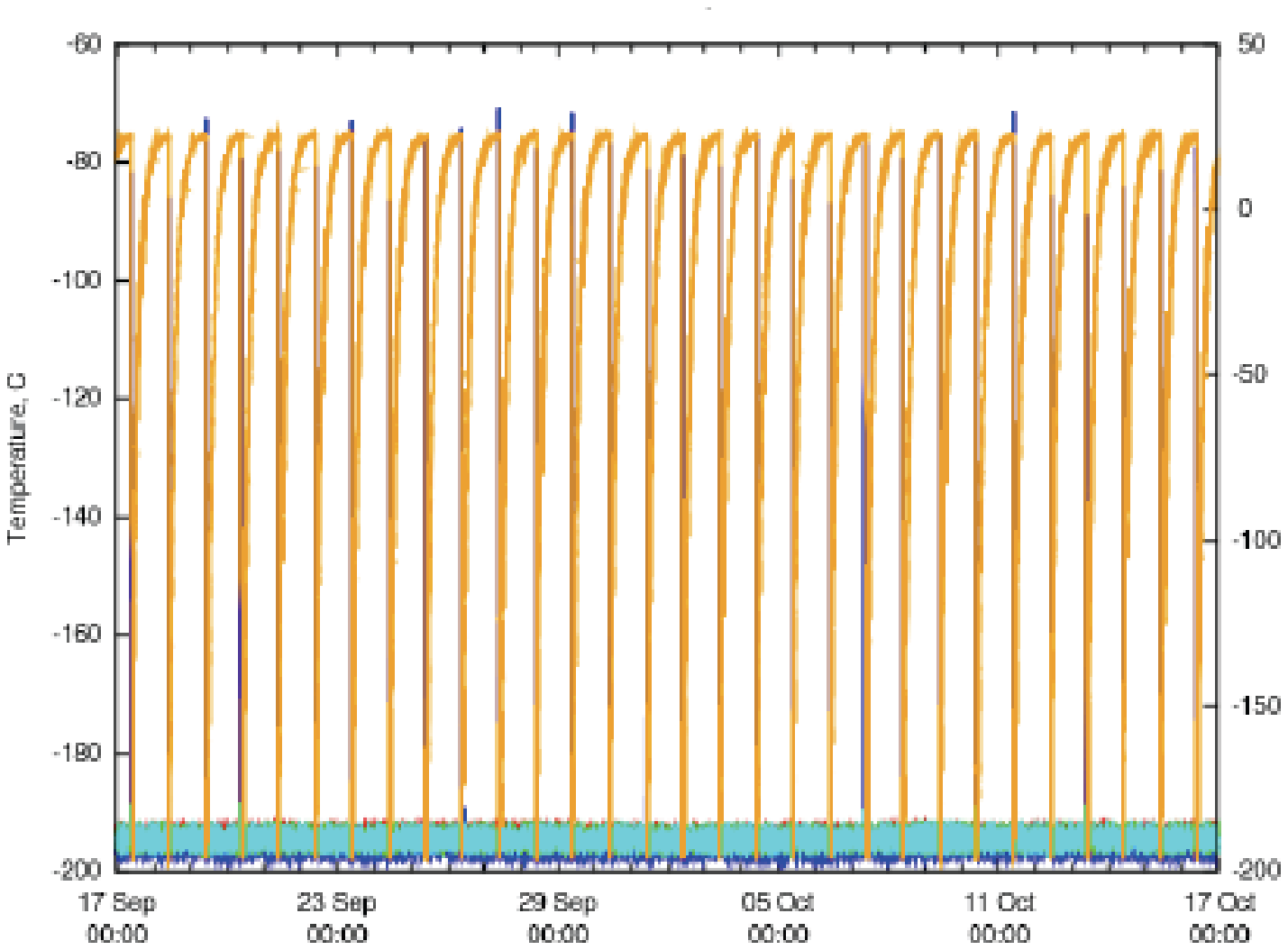}\includegraphics[width=.42\textwidth]{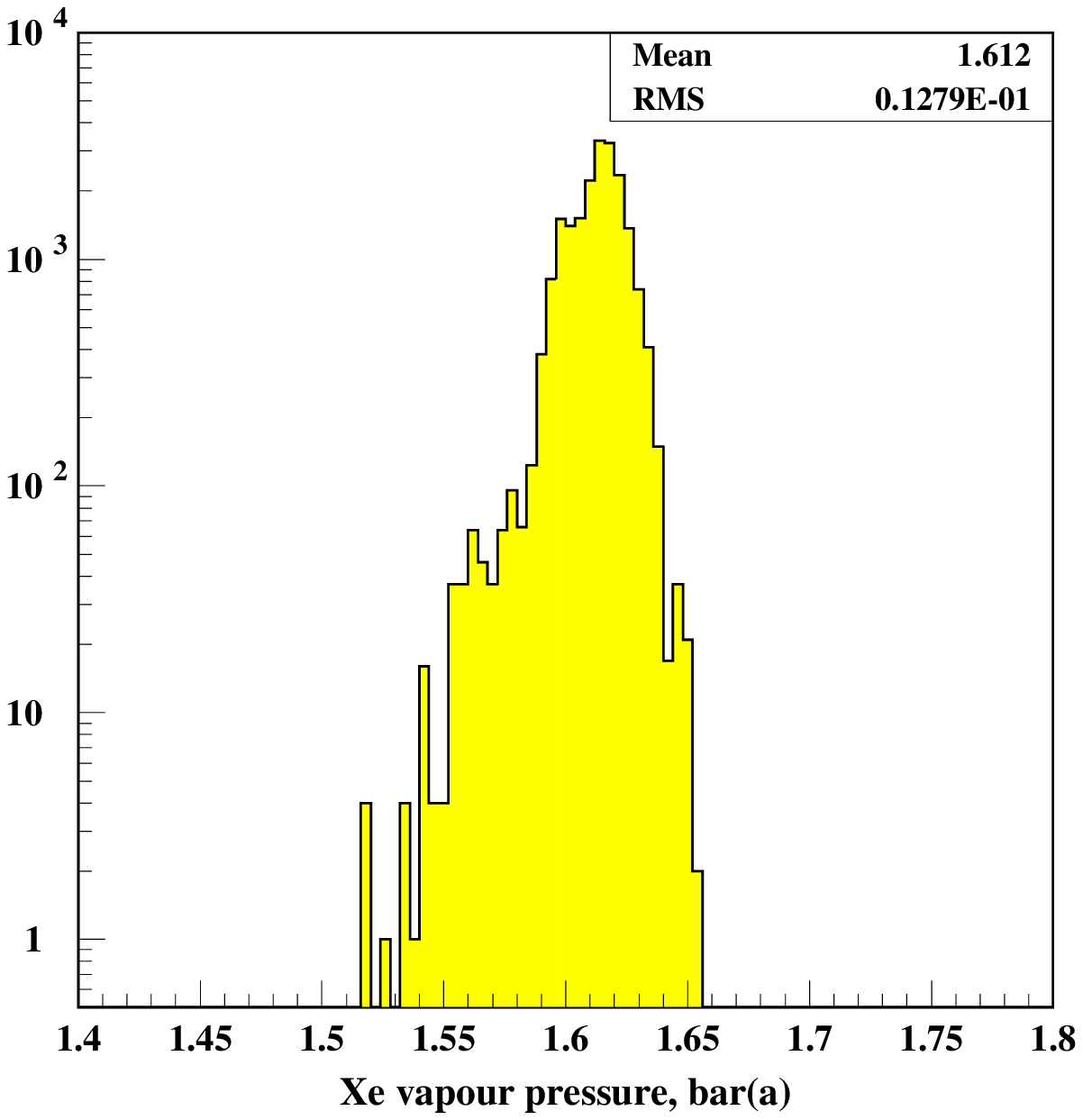}
\caption{Left: History of one month's worth temperature at the end of the liquid nitrogen delivery line showing an excellent reproducibility achieved 
thanks to the automation of the detector servicing process. Right: Distribution of the vapour pressure above the liquid xenon target with RMS 
below 1\% of the mean.}  
\label{fig3}
\end{figure}

Daily detector operations -- including detector calibration, liquid nitrogen (LN$_2$) refill and the data transfer -- interrupted the science 
run for only one hour, thus achieving routinely a 96\% duty cycle. An example of the excellent stability and reproducibility is shown in Figure~\ref{fig3}, plotting  
one month's worth of daily LN$_2$ refills which occured every day exactly at the same time. Pressure of the xenon vapour above the target was kept 
at 1.6 bar, with a rms variability at the level of 1\%. 

Slow movement of the rock underneath the detector, causing a tilt, was monitored using the width of the S2 
signal from the calibration data. The history of the correction factor due to detector tilt is shown in Figure~\ref{fig4}. The tilt was 
rectified weekly to first order using the pulley system visible in Figure~\ref{fig2}. Long exposure $\gamma$ and 
neutron calibration runs were carried out at the beginning and at the end of the data taking run. To control PMT performance, weekly calibrations were 
performed with an LED gun coupled to a quartz optical fibre which delivered light directly into the xenon target.

\subsection{Data acquisition and processing}

Since the PMT array was powered by a single high voltage supply all signal outputs were equalized with a set of Phillips Scientific 804 
attenuators. Subsequently, the signals were split into two channels and digitized with 2 ns sampling using ACQIRIS DC256 8-bit flash ADCs.
In one of the channels signals were amplified $\times$ 10 by fast Phillips Scientific 770 amplifiers to achieve a higher dynamic range.
Additionally all signals were summed with a Hoshin N005 sum amplifier with the output signal used to trigger the DAQ system.

The acquired 36 $\mu$s long waveforms were reduced with ZE3RA~\cite{ze3ra} software, producing an array of parameters 
for the 10 largest pulses. Afterwards an event filtering tool was used to retain events containing only one pair of pulses coming from the liquid scintillation (S1) and the gas electroluminescence (S2). The energy and position reconstruction, from the S1 and S2 signals, for each event was performed with the bespoke software 
Mercury~\cite{merc} using pulse areas and the {\it z} coordinate calculated from S2-S1 timing information. 

\begin{figure}[t]
\includegraphics[width=.405\textwidth]{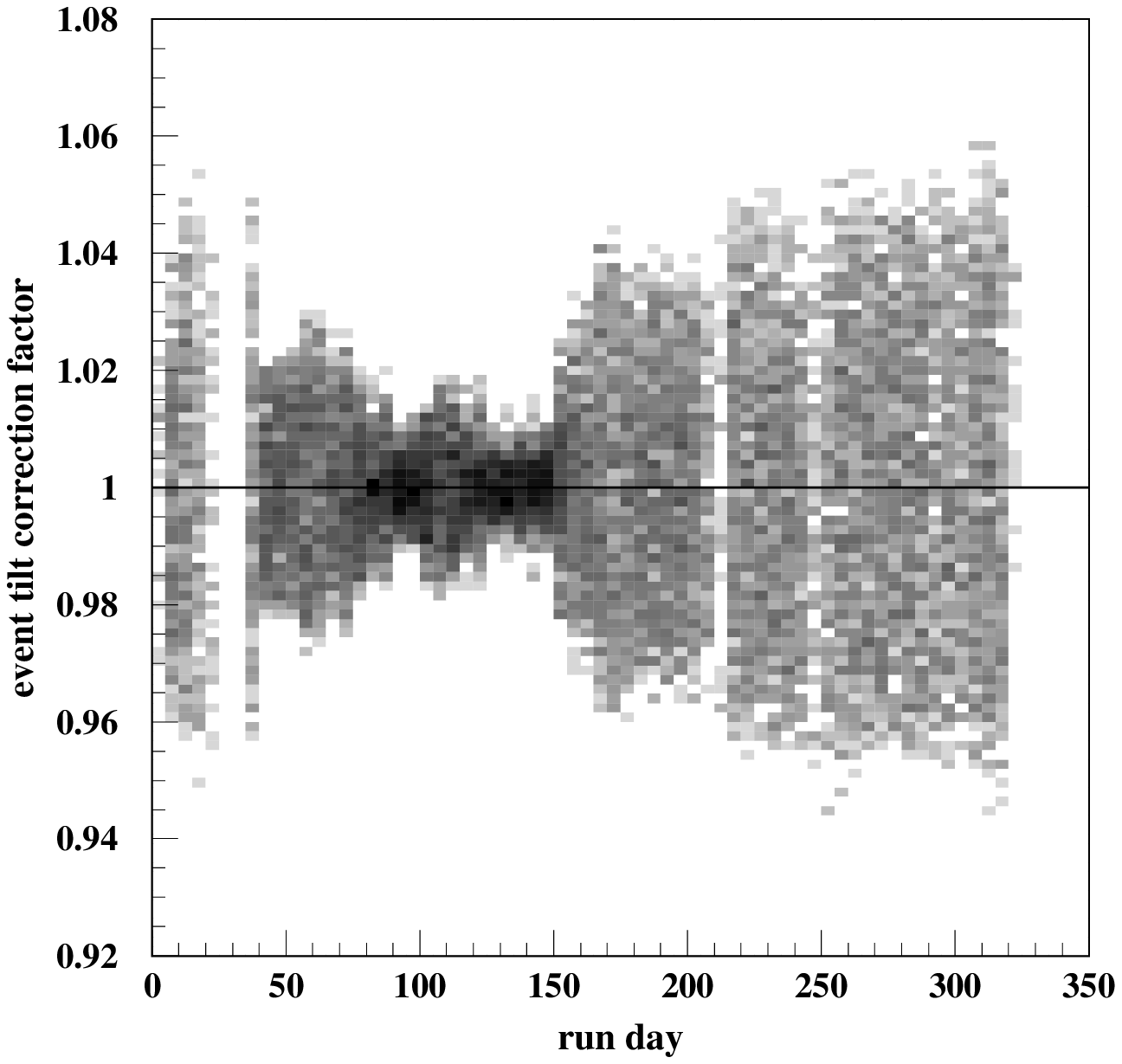}\includegraphics[width=.595\textwidth]{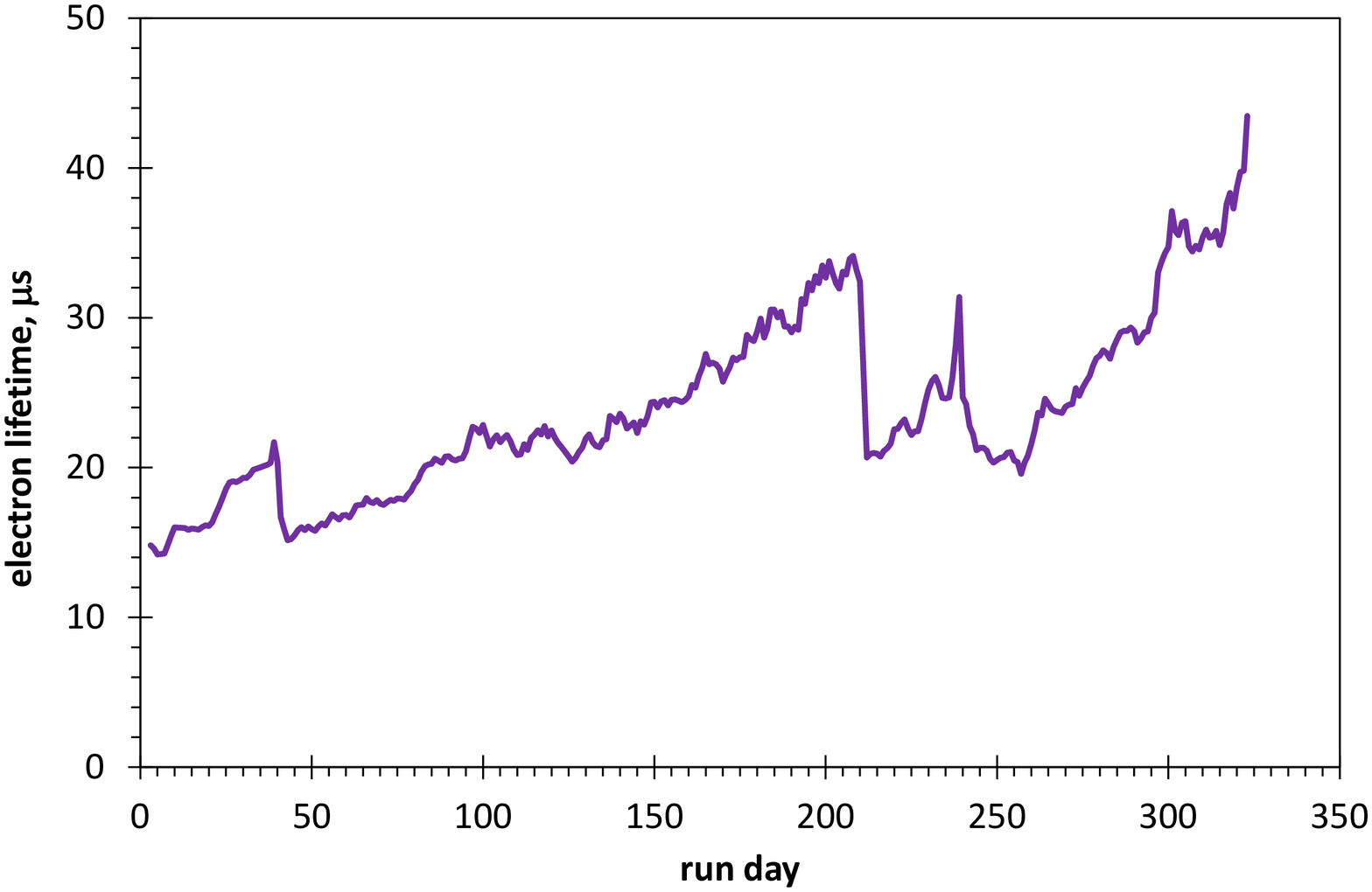}
\caption{Left: Distribution of correction factors for S2 pulse area due to geology-induced detector tilt in the second run. The tilt is mitigated 
to first order with the pulley system visible in Figure 2 and the residual variation is corrected in software. This parameter is calculated from 
the variation of S2 pulse width across the vapour phase in the calibration data and then applied to the $x,y$ position of each event.
Right: Evolution of the free electron lifetime in the second run.}
\label{fig4}
\end{figure}

In addition to the data processing the following corrections have been applied to each selected single scatter event: electronics gain drift, 
detector tilt, vapour pressure variation and electron lifetime in the liquid which, as shown in Figure~\ref{fig4}, with an initial value of 14 $\mu$s at the beginning of the run gradually 
increasing to 45 $\mu$s by the end of the run. Although there is no gas recirculation/purification once the detector is in operation, this parameter 
improved steadily during the run. This is mainly due to the sweeping of electronegative ions away from the LXe bulk by the electric field and, 
to a smaller extent, due to gettering of electronegative impurities by the detector components. Periods of sharp lifetime degradation correspond 
to power failures at the underground lab. Finally, all events were checked for veto tagging of any kind. A total of 28\% of 
all $\gamma$-ray events were flagged as PTAG as expected for the $\gamma$-ray background.

\subsection{Event energy and position reconstruction}  

Event reconstruction consists of estimating the energy and the position of an event given a set of the corresponding PMT pulse areas. 
For an event at position $r$ producing $N$ photons, the probability of the i-th PMT detecting $n$ photons is well approximated by 
the Poisson distribution:
\begin{equation}
P_i(n)=\frac{\mu_i^{n}e^{-\mu_i}}{n!}\qquad,
\label{eq1}
\end{equation}
where $\mu_i$=N$\eta_i$($r$) is the expected number of photons from  $N$ initial photons detected by the i-th PMT with the 
$\eta_i$($r$) being the Light Response Function (LRF) -- the fraction of the photons emitted by the source that produce a detectable signal 
in the i-th PMT.

In this case the interaction location ($r$) and the total number of emitted photons ($N$) can be found by the Maximum Likelihood (ML)
method as was first proposed by Gray and Makovsky~\cite{gm}. Given the number $n_i$ of photons detected by each PMT, the logarithm of the likelihood 
function can be expressed as~\cite{cook}:
\begin{equation}
\ln L(\textbf{r},N)=\sum_i(n_i\ln(N\eta_i(\textbf{r}))-N\eta_i(\textbf{r}))+C\qquad,
\label{eq2}
\end{equation}
where $C$ depends only on the n$_i$.

Different statistical approaches have been applied to reconstruct the S1 and S2 signals. Since the total collected charge in the S1 signal equals only 1.2 phe/keV 
the statistical variation of the number of photoelectron in each PMT is Poissonian; in this case the event is reconstructed by maximizing the above function. On 
the other hand, the size of the S2 pulse is boosted by the electroluminescence 
in the gas (a single electron extracted to the gas phase produces an average of 12 photoelectrons in the SSR configuration~\cite{sep}); this is typically 
two orders of magnitude larger than S1. In this instance the reconstruction is performed with a weighted least square (WLS) method. In this case the parameters 
were obtained through the following minimisation: 
\begin{equation}
\chi^2=\sum_{i}{w_i(A_{ei}-A_i)^2};  A_{ei}=N\eta_i(\textbf{r})q_{si}\qquad,
\label{eq3}
\end{equation}
where $A_i$ and $A_{ei}$ are the measured and the expected output signals of the i-th PMT, respectively; $q_{si}$ is the mean of the single photoelectron probability 
density function (PDF) and $w_{i}$ is the weighting factor related to the variance of $(A_{ei}-A_i)$. 
 
Both the ML and WLS methods can be used only if the LRF is well known. In principle, it should be either measured or calculated beforehand. In the 
case of ZEPLIN-III, however, the direct measurement poses great technical difficulty and calculation could not provide the necessary precision. For 
this reason, a novel method was developed for {\it in-situ} reconstruction of the PMT light response functions (LRF). These were obtained from the calibration data 
acquired by irradiating the detector with an uncollimated $\gamma$-ray source. The method takes advantage of the fact that, for the ZEPLIN-III geometry, the LRFs
are functions of the distance from the PMT axis only. These LRFs were reconstructed iteratively from a set of 122 keV events from the $^{57}$Co calibration. 

\begin{figure}[ht]
\includegraphics[width=.465\textwidth]{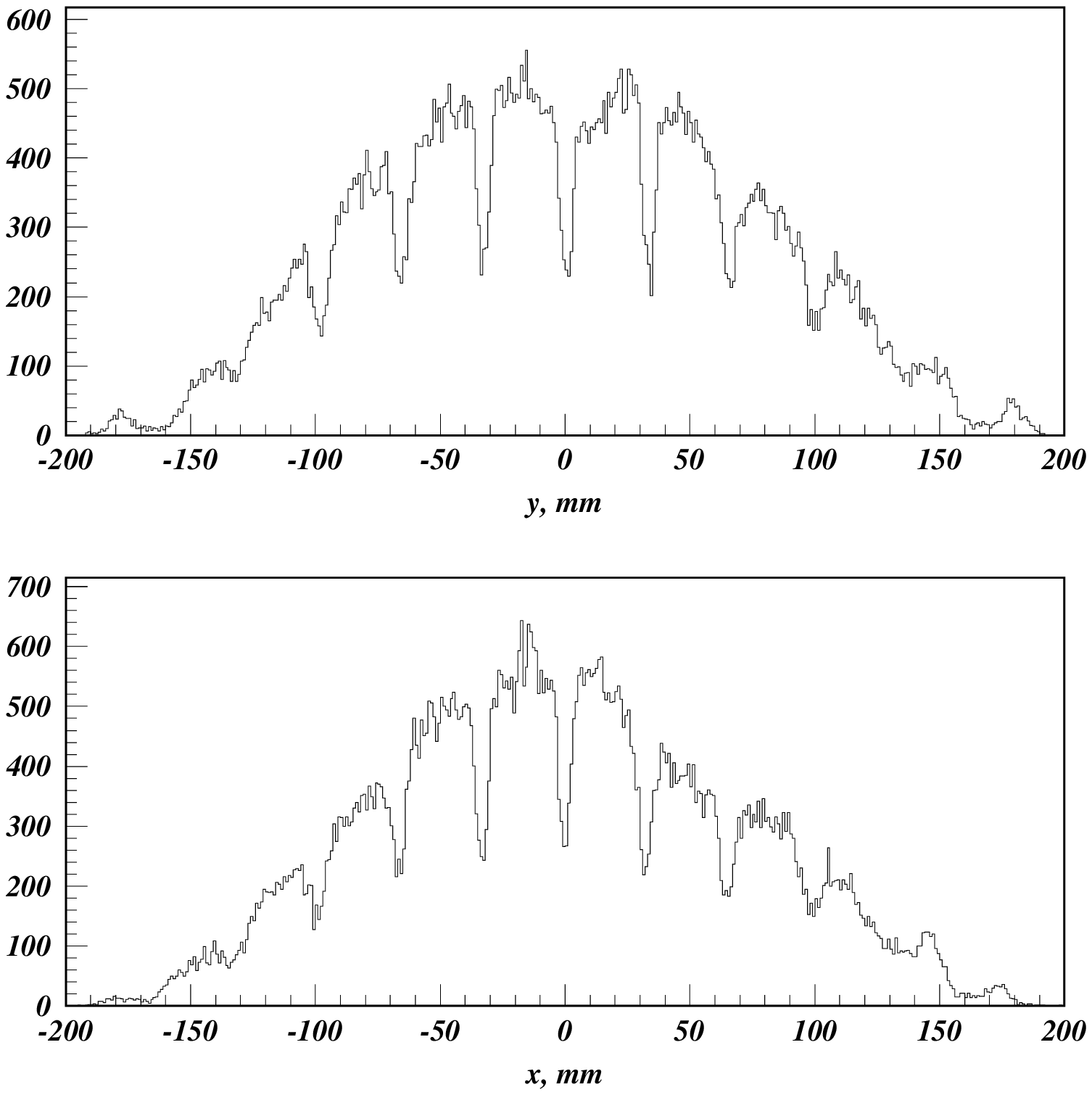}\includegraphics[width=.535\textwidth]{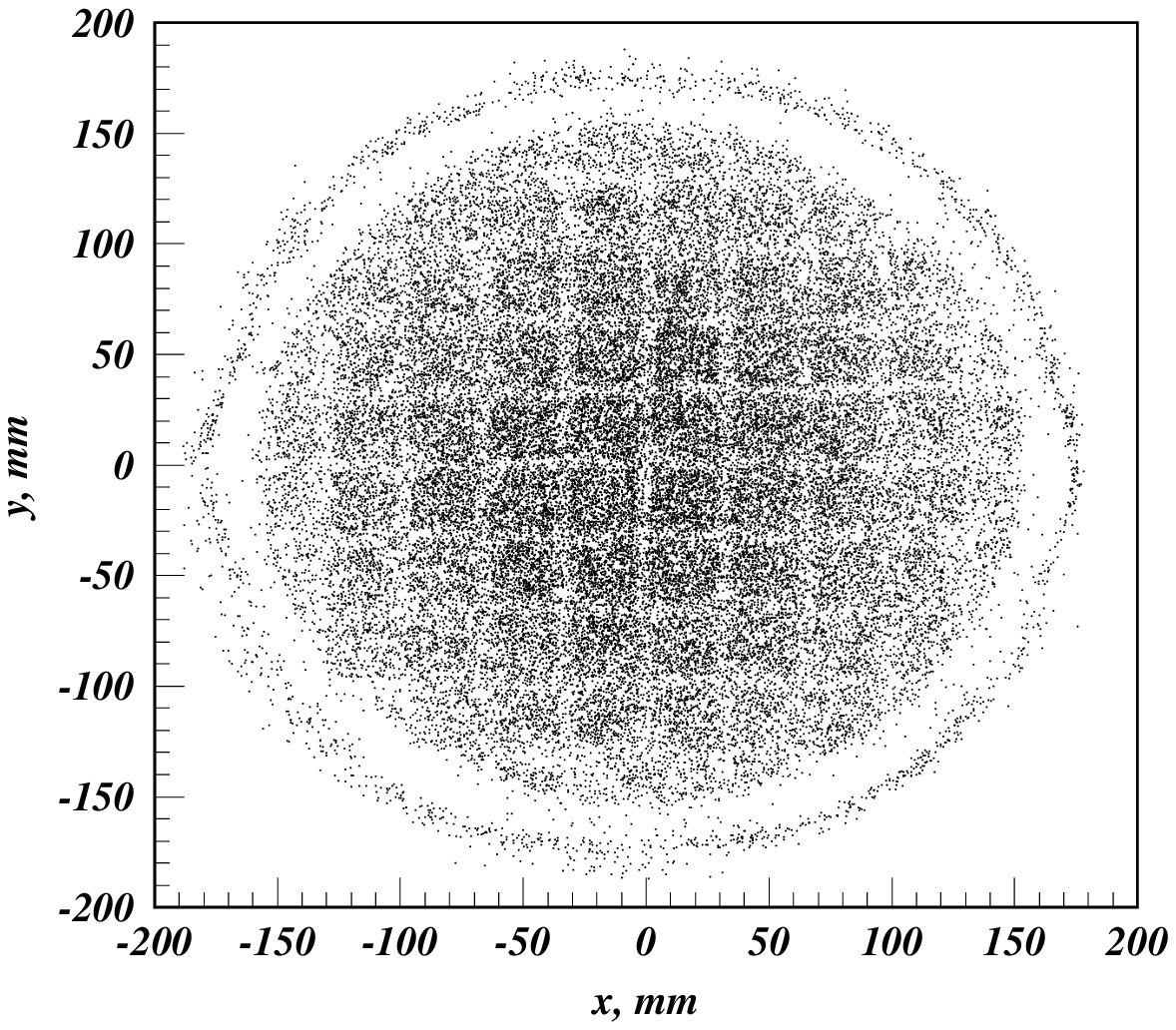}
\caption{Left: 122 keV $\gamma$ vertex profile from an external source placed above the phantom grid. Right: S2 signal X-Y event reconstruction 
using WLS method showing a very clear image of the phantom grid.}  
\label{fig5}
\end{figure}

In the first step, the $x,y$ vertex positions were estimated by means of a simple centroid algorithm. For each PMT the area 
of S2 pulses was plotted versus reconstructed distance and fitted by a smooth non-increasing function; this then became the first approximation for the corresponding 
LRF. Using this first approximation, the event positions were re-estimated by the WLS method and the fitting was repeated giving the next (better) approximation
for the LRFs. The iteration continued until the response function converged. A reconstructed image of the phantom grid after five iterations is shown 
in Figure~\ref{fig5}. On the same figure, the profiles of the reconstructed event density are shown for the $x$ and $y$ directions. From these profiles, a 
spatial resolution of 1.1 mm (FWHM) was calculated for S2 signals. The spatial resolution for S1 was measured as the spread of the S1 position with respect to that 
of S2 and was estimated to be 13 mm. Note that the sharpness of the image is dominated by scattering of the $\gamma$-rays in the 7-mm thick anode mirror under the 
copper grid, rather than by the position resolution. 

A linear combination of S1 and S2 signals from $^{57}$Co source was used to estimate the energy resolution at 122 keV. In the central region of 50 mm radius 
the reconstructed energy resolution for the FSR and the SSR were 8.1\% and 12\%, respectively. In case of the FSR dataset the line at 136 keV 
was clearly resolved.

\section {WIMP Results}
 
\begin{figure}[ht]
\includegraphics[width=.523\textwidth]{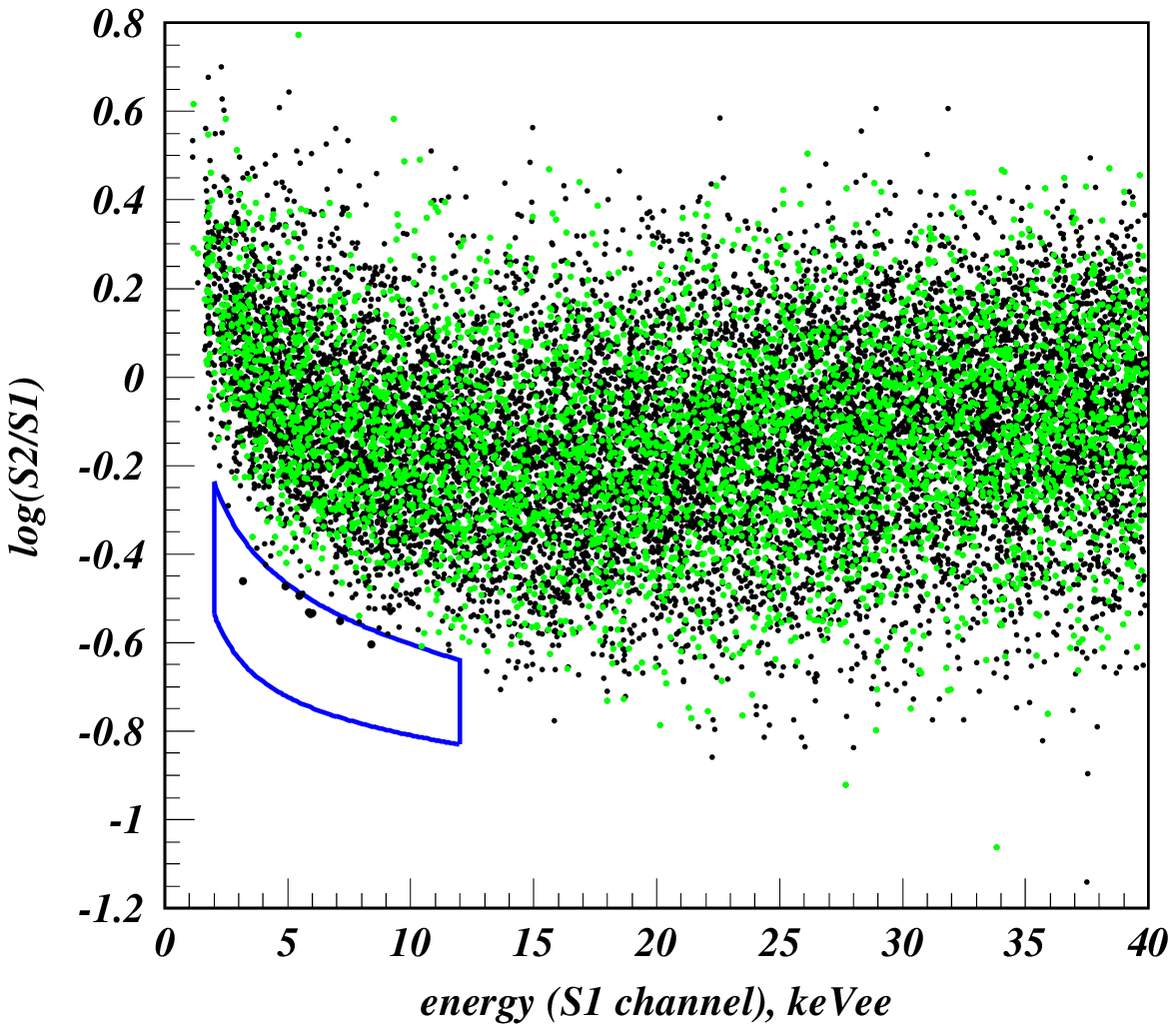}\includegraphics[width=.477\textwidth]{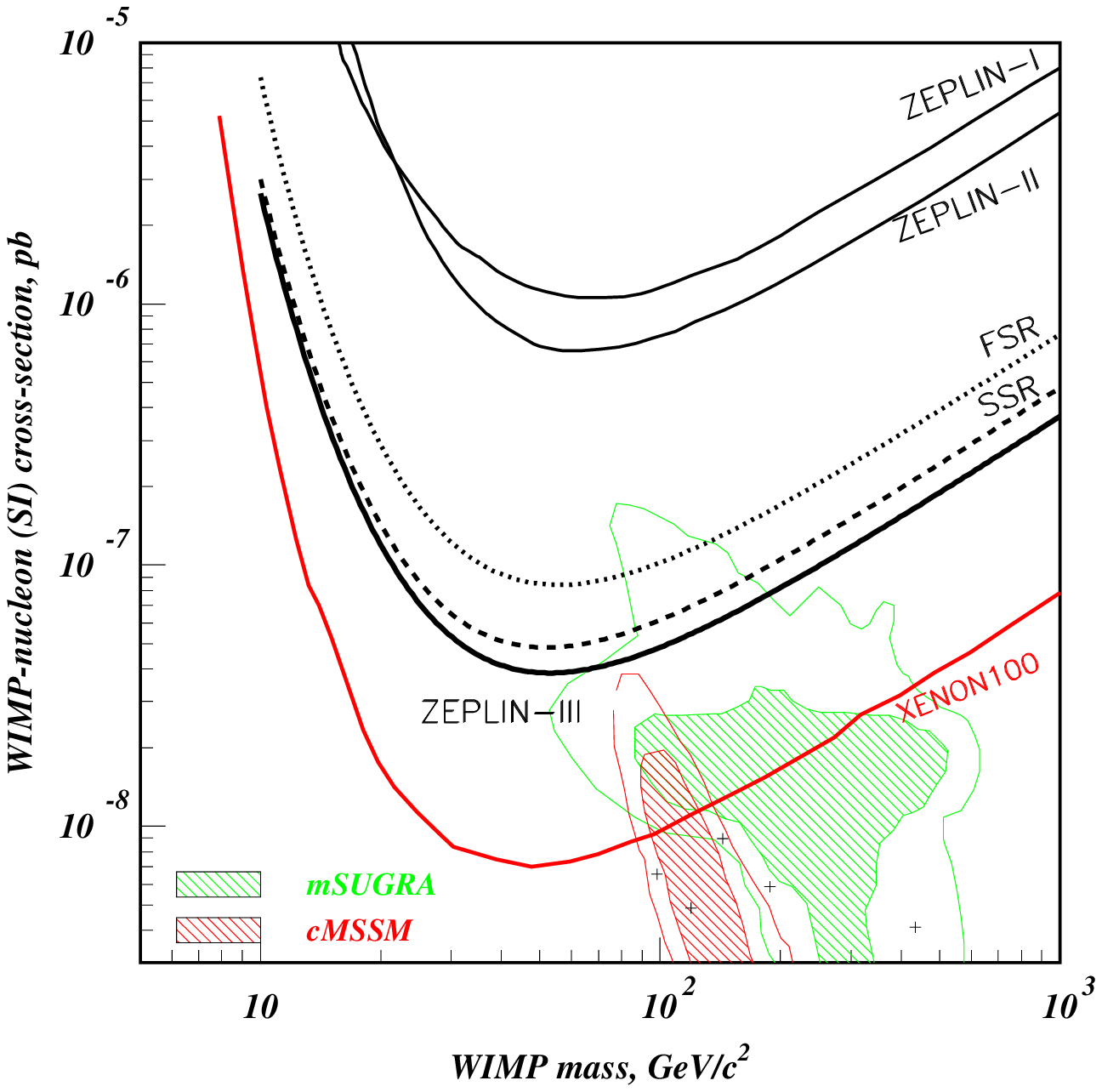}
\caption{Left: S2/S1 distribution as a function of energy of SSR events in the fiducial region. Events marked with a green color represent PTAG veto coincidences. 
8 unvetoed points are in the WIMP search box (blue line). Right: 90\% CL limits on WIMP-nucleon scalar cross sections from all ZEPLIN programme experiments as well as from the 
leading XENON-100 experiment.}
\label{fig6}
\end{figure}

The fiducial region of the detector was chosen to be a central cylinder of 140 mm radius and contained 5.1 kg of liquid xenon. 
Figure~\ref{fig6} shows all events in the S2/S1 parameter space from the final analysis reported in~\cite{ssr}. The acceptance region for the WIMP 
search was defined between 2 and 12 keV electron-equivalent energy and to contain 2 -- 45\% acceptance in the log(S2/S1) parameter. 
This is below the mean of the nuclear recoil band derived from the neutron calibration~\cite{horn}. Eight events were found in the box, none of which had been vetoed. 
The number of DTAG events in the dataset was consistent with random coincidences 
and none were below the nuclear recoil median. The number of nuclear recoils events predicted for the search region was 0.06$\pm$0.01. The number of electron 
recoils events leaking into the WIMP search box was estimated in two ways, from a dedicated $^{137}$Cs calibration run and using binned skew-Gaussian 
fits to the gamma band above the search box, giving 9.3$\pm$3.9 and 6.5$\pm$3.4 events, respectively. 

A binned profile likelihood ratio~\cite{plr1,plr2} statistical analysis yielded a two-sided confidence interval 0--5.1 signal events at 90\% CL as described in detail 
in ~\cite{ssr}. New upper limits on the WIMP-nucleon scalar cross-sections derived for the FSR, SSR and the combined exposure are presented in Figure~\ref{fig5}.
To show the progress of the entire ZEPLIN programme at Boulby, which produced competitive results for over a decade, both limit results from ZEPLIN-I 
and ZEPLIN-II have been also included in Figure~\ref{fig5}. The present best world limit by XENON100~\cite{aprile}, another 
two-phase xenon experiment, is also shown. 

\section{Conclusions}
ZEPLIN-III was upgraded with new photomultipliers and an anti-coincidence detector was also installed. The latter served not only as a neutron detecting 
system but also as an excellent diagnostic tool for the $\gamma$-ray background. The prediction of 28\% of PTAG events was confirmed showing 
a very good understanding of the dominant radiation background. Thanks to the newly automated daily detector 
operations including $\gamma$-ray calibrations, LN$_2$ refill and the data transfer a 96\% duty cycle was achieved routinely. An excellent reproducibility,
and control of detector parameters such as: liquid purity, detector tilt, gas gap thickness and the xenon vapor pressure contributed greatly to the very competitive 
final results from the SSR. Achieving good vertex and energy reconstruction with poorly performing photomultipliers was a key challenge in data analysis.  
The second science run delivered a 90\% CL upper limit on the scalar WIMP cross-section of 4.8$\times$10$^{-8}$ pb/nucleon near 50 GeV/c$^2$ mass. The combined 
result from the FSR and SSR is 3.9$\times$10$^{-8}$ pb/nucleon. 

The instrument performance in 319 days of the second run demonstrated clearly that xenon emission detectors can possess the required long-term stability and reliability 
for rare event searches (even if LN$_2$-cooled). 

\section{Acknowledgments}
The UK groups acknowledge the support of the Science \& Technology Facilities Council (STFC) for the ZEPLIN\textendash{}III project and
for maintenance and operation of the underground Palmer laboratory which is hosted by Cleveland Potash Ltd (CPL) at Boulby Mine, near
Whitby on the North-East coast of England. The project would not be possible without the co-operation of the management and staff of CPL.
We also acknowledge support from a Joint International Project award, held at ITEP and Imperial College, from the Russian Foundation of
Basic Research (08-02-91851 KO a) and the Royal Society. LIP\textendash{}Coimbra acknowledges financial support 
from Funda\c{c}\~ao para a Ci\^encia e a Tecnologia (FCT) through the project-grants CERN/FP/109320/2009 and CERN/FP/116374/2010,
as well as the postdoctoral grants SFRH/BPD/27054/2006, 
\newline SFRH/BPD/47320/2008 and SFRH/BPD/63096/2009. This work was supported in part by SC Rosatom,
contract \#H.4e.45.90.11.1059 from 10.03.2011.

\end{document}